# II-VI Organic-Inorganic Hybrid Nanostructures with Greatly Enhanced Optoelectronic Properties, Perfectly Ordered Structures, and Over 15-Year Shelf Stability

*Tang Ye[1], Margaret Kocherga[1], Yi-Yang Sun[2], Andrei Nesmelov[3], Fan Zhang[4], Wanseok Oh[4], Xiao-Ying Huang[5,6], Jing Li[5], Damian Beasock[1], Daniel S. Jones[3], Thomas A. Schmedake[1,3], and Yong Zhang[1,4]\**

[1]Nanoscale Science, University of North Carolina at Charlotte, Charlotte, NC 28223, USA.

[2]State Key Laboratory of High-Performance Ceramics and Superfine Microstructure, Shanghai Institute of Ceramics, Chinese Academy of Sciences, Shanghai 201899, China.

[3]Department of Chemistry, University of North Carolina at Charlotte, Charlotte, NC 28223, USA.

[4]Department of Electrical and Computer Engineering, University of North Carolina at Charlotte, Charlotte, NC 28223, USA.

[5]Department of Chemistry and Chemical Biology, Rutgers University, Piscataway, NJ 08854, USA.

[6]State Key Laboratory of Structural Chemistry, Fujian Institute of Research on the Structure of Matter, Chinese Academy of Sciences, Fuzhou, Fujian 350002, P. R. China.

*Correspondence: yong.zhang@uncc.edu

This is the final version accepted by ACS Nano.

**Abstract**

Organic-inorganic hybrids may offer material properties not available from their inorganic components. However, they are typically less stable and disordered. Long-term stability study of the hybrid materials, over the anticipated lifespan of a real-world electronic device, is practically nonexistent. Disordering, prevalent in most nanostructure assemblies, is a prominent adversary to quantum coherence. A family of perfectly ordered II-VI based hybrid nanostructures has been shown to possess a number of unusual properties and potential applications. Here, using a prototype structure β-ZnTe(en)$_{0.5}$ – a hybrid superlattice, and applying an array of optical, structural, surface, thermal, and electrical characterization techniques in conjunction with density-functional theory calculations, we have performed a comprehensive and correlative study of the crystalline quality, structural degradation, electronic, optical, and transport properties on samples from over 15-years old to the recently synthesized. The findings show that not only do they exhibit an exceptionally high level of crystallinity in both macroscopic and microscopic scale, comparable to high-quality binary semiconductors; and greatly enhanced material properties, compared to those of the inorganic constituents; but also, some of them over 15-years old remain as good in structure and property as freshly made ones. This study reveals (1) what level of structural perfectness is achievable in a complex organic-inorganic hybrid structure or a man-made superlattice, suggesting a non-traditional strategy to make periodically stacked heterostructures with abrupt interfaces; and (2) how the stability of a hybrid material is affected differently by its intrinsic attributes, primarily formation energy, and extrinsic factors, such as surface and defects. By correlating the rarely found long-term stability with the calculated relatively large formation energy of β-ZnTe(en)$_{0.5}$ and contrasting with the case of hybrid perovskite, this work illustrates that formation energy can serve as an effective screening parameter for the long-term stability potential of hybrid materials. The results of the prototype II-VI hybrid structures will on one hand inspire directions for future exploration of the hybrid materials, and on the other hand provide metrics for assessing the structural perfectness and long-term stability of the hybrid materials.

Organic-inorganic hybrid materials have been explored for a wide range of applications because of their added and enhanced properties compared to their inorganic counterparts.[1-6] Among them, hybrid halide perovskites (*e.g.*, MAPbI$_3$, MA = CH$_3$NH$_3$) are perhaps the most extensively and intensively studied hybrid materials, because of their extraordinary application potentials in a wide range of applications, in particular photovoltaics (PV) and solid-state lighting (SSL). However, hybrid materials typically exhibit two major drawbacks: (1) lower long-term stability than inorganic compounds; and (2) structural disorder. These two characteristics make them distinctly different from typical crystalline inorganic semiconductors and related nanostructures.

Systematic long-term stability studies of hybrid materials in general are understandably rare, if any, since they typically do not last long. By long-term, we mean a time scale of one or two decades that is typically expected for (opto-)electronic applications. Long-term stability and high melting points of familiar semiconductors like Si and GaAs are intrinsically associated with their large formation energies. However, extrinsic effects, such as surface and defects, often affect their stability under ambient condition. Luckily, for many crystalline inorganic materials, the oxidation is usually a self-limiting process, which ensures their long-term stability. Hybrid perovskites are known to have relatively poor long-term stability, both under ambient conditions[6] and illumination,[7] which limits the scope of their applications. Despite the major improvement in stability from hours just a few years ago[7] to currently up to a couple of months, benefiting from reduction in structural defects and surface passivation,[8-10] and some 2D versions show moderately better stability,[11, 12] they remain inadequate for general applications of PV and SSL. The degradation mechanisms are not yet well understood. It has been shown theoretically that the hybrid perovskite MAPbI$_3$ has very low or even negative formation energy (±0.1 eV).[13, 14]

Would this place a limit on how much further improvement is attainable? There is neither a reliable theoretical method to predict the long-term stability of a hybrid material nor a hybrid material with demonstrated long-term stability that can serve as an example to show the necessary characteristics. Can indeed a hybrid with a larger formation energy offer long-term stability? How would the roles of the extrinsic effects be in that case? Therefore, investigating the long-term evolution of a hybrid material that does have good long-term stability becomes critically important and highly appreciated.

Hybrid perovskites are considered as crystalline materials in the sense that they have excellent long-range order. Nevertheless, they are structurally disordered in short-range in their room temperature phases,[15, 16] thus, not crystals in the genuine sense. One might think that inorganic nanostructures, the simplest cases – semiconductor superlattices, are easier to achieve perfect crystal structures. In reality, although it has been over five decades since the proposal of the superlattice concept,[17] the holy grail of the superlattice, a structure without inter-diffusion between the adjacent monolayers of the constituents, remains elusive, even for the best studied examples $(GaAs)_n/(AlAs)_m$ and $(GaP)_n/(InP)_m$.[18-20] Although disordering can be beneficial for some applications,[16, 21] it diminishes quantum coherence, which is required for various advanced applications, such as quantum optics,[22] and affects negatively on many basic material properties, such as exciton dynamics and electronic conductivity.[23] Therefore, a synthesis approach that can yield a perfectly ordered nanostructure array will be of both fundamental and practical interest.

The II-VI based hybrid nanostructures were first reported in 2000,[24] including a large number of combinations of II-VI compounds and molecules as well as distinctly different structure types.[25] As a prototype structure, β-$ZnTe(en)_{0.5}$ (en = $C_2N_2H_8$, ethylenediamine) consists of two-monolayer thick ZnTe (110) slabs interconnected by en molecules forming covalent-like bonding through Zn and N atoms,[24] as shown in **Figure 1**a. Although it is a crystal on its own with sub-nm scale integration of the organic and inorganic components, it can nevertheless be viewed as a hybrid superlattice with the molecule serving as electronic barriers as well as dielectric confinement layers.[26] In contrast to the inorganic superlattices and most other hybrid materials, this hybrid superlattice is a perfectly ordered structure microscopically. It has both long and short-range order,[24] because the organic molecules all stay in well-defined configurations, and the inter-diffusion between the inorganic and organic layers is not possible due to their vastly different structures. β-ZnTe(en)$_{0.5}$ has been shown to have crystallinity comparable to most high-quality elemental and binary semiconductors, *e.g.*, low-temperature Raman linewidth below 1 cm$^{-1}$.[25, 27] More interestingly, it exhibits various desirable properties, *e.g.*, room temperature excitonic emission due to a large exciton binding energy estimated to be over 200 meV, strongly enhanced optical absorption as high as $10^6$ cm$^{-1}$,[26] broad-range zero-thermal expansion,[27] much-reduced density and dielectric constants,[26] and greatly reduced thermal conductivity.[28] These properties suggest a number of potential applications, including room-temperature exciton-polariton condensation, optical switching, efficient UV emission and detection, and thermoelectrics. It has recently been demonstrated that one could even exfoliate the thin inorganic slabs individually from these II-VI hybrids.[29] These attributes make β-ZnTe(en)$_{0.5}$ an ideal prototype system for revealing the pertinent mechanisms dictating the long-term stability of hybrid materials and general design principles to achieve highly ordered hybrid structures. This work conducts a systematic investigation for β-ZnTe(en)$_{0.5}$ to reveal how practically perfect structural ordering, in both a macroscopic and microscopic sense, is manifested in its structural and electronic characteristics, and how the intrinsic and extrinsic mechanisms affect its structural and physical properties over one and a half decades.

## Results and Discussion

### Structural ordering, optical properties, and stability probed by optical spectroscopy

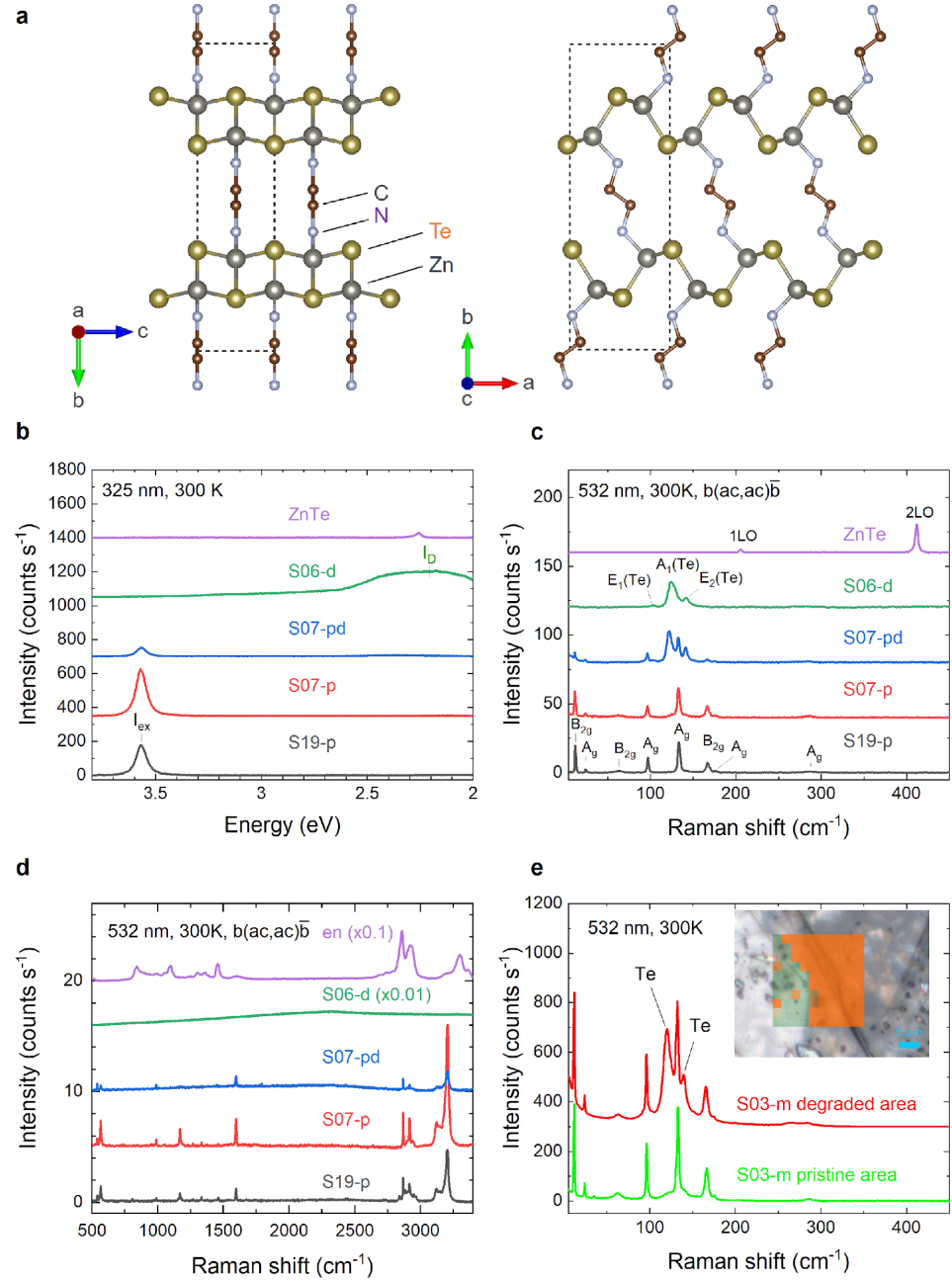

FIG. 1. Optical spectroscopy signatures of β-ZnTe(en)$_{0.5}$ at different degradation stages. a) Crystal structure viewed along a-axis and c-axis. b) Photoluminescence spectra. c) Raman spectra of low wavenumber region. d) Raman spectra of large wavenumber region. e) Raman spectra of pristine and partially degraded areas adjacent to each other (inset: Raman mapping of the Te mode showing no sign of degradation in the green area but partial degradation in the red area).

β-ZnTe(en)$_{0.5}$ is an orthorhombic crystal with space group *Pnnm*, point group D$_{2h}$, and stacking axis *b* and in-plane axes *a* and *c*, typically found as a thin crystal plate with the *b* axis aligned with the normal.[24, 26] Pristine β-ZnTe(en)$_{0.5}$ is a transparent crystal with a bandgap 3.719 eV at 1.5 K.[26] It exhibits distinctly different optical properties from those of ZnTe. **Figure 1**b–d, respectively, compare the spectra of photoluminescence (PL) excited at 325 nm and Raman excited at 532 nm (in small and large frequency regions) of the hybrid samples in pristine and varying degree of degradation states, with comparison to those of ZnTe. In **Figure 1**b, the PL spectrum of a fresh-made sample S19-p ("19" stands for the year of synthesis 2019, "p" for pristine) shows only a single peak ($I_{ex}$) at 3.569 eV of free exciton emission.[26] This hybrid crystal is free of below bandgap defect emission, which is in stark contrast to most high bandgap inorganic semiconductors such as GaN and ZnO, where below bandgap defect emission is almost inevitable.[30, 31] In **Figure 1**c, the Raman spectrum of S19-p, measured in a backscattering geometry along the *b* axis and parallel polarization configuration of 45° between *a* and *c* axes to reveal a maximum number of Raman modes intrinsic to the hybrid structure, exhibits multiple sharp peaks similar to those reported previously[27] but with two additional lower frequency modes (see Table S1 for Raman mode frequencies and symmetries). The symmetry assignments were obtained by DFT calculations of phonon modes and their Raman tensors. Despite being a much more complex structure (32 atoms/unit cell), the Raman linewidths (measured by full width at half maximum, FWHM) of the hybrid is comparable to that of ZnTe: *e.g.*, 3.2 *vs.* 3.7 cm$^{-1}$ between the strongest hybrid mode at 133.2 cm$^{-1}$ and ZnTe 1LO at 205.8 cm$^{-1}$, indicating a high degree of structural ordering in the hybrid. The Raman spectrum of S19-p in the higher frequency region is shown in **Figure 1**d. Because of the strong bonding between the ZnTe sheets and en molecules, the ordering also reflects on the en derived Raman modes, where the en modes (see Table S1 for details) are typically shifted in frequency but substantially narrower, compared to those of free-standing en,[32] for instance, 2869 (FWHM = 5) *vs.* 2860 (FWHM = 36) cm$^{-1}$. It is worth noting that β-ZnTe(en)$_{0.5}$ studied here has less Raman modes in the high frequency region than α-ZnS(en)$_{0.5}$ measured previously[33] because the α phase has twice as many atoms in the unit cell.

A number of aged samples were measured under the same conditions. Quite remarkably, the same intrinsic PL and Raman spectra were observed in some samples that were more than a decade old, although the aged samples often exhibited varying degrees of degradation, manifested as changes in both PL and Raman spectrum. They fall into three groups: (1) pristine: showing practically identical optical spectra as S19-p, (2) degraded: showing no intrinsic hybrid spectroscopy features but those of the degradation products, (3) partially degraded: showing the spectroscopy features of both (1) and (2). Representative results are shown in **Figure. 1**. For example, S07-p, typical from a batch of crystals, exhibits nearly identical spectroscopy signatures of S19-p, as shown in **Figure 1**b for PL, **Figure 1**c and **1**d for Raman. Typical PL and Raman spectra for a partially degraded sample (S07-pd) and severely degraded sample (S06-d) are shown in **Figure 1**b, **1**c and **1**d, respectively. They show in PL the weakening or disappearance of the band edge emission peak and appearance of a broad below bandgap emission band (I$_D$), whereas in Raman the weakening or disappearance of the hybrid Raman modes, and most noticeably the appearance of Raman modes related to metallic Te at ~124 and 142 cm$^{-1}$.[34] Degradation is often non-uniform over the surface area of a crystal. For instance, the intrinsic Raman features were observed in some even older samples, such as S03-m ("m" for mixed states of degradation), where adjacent pristine and partially degraded regions co-existed, as shown in **Figure 1**e. Similar spatial variation was also observed in a bulk hybrid perovskite crystal,[7] but the overall degradation occurred in a totally different time scale. Although β-ZnTe(en)$_{0.5}$ is still not as stable as inorganic semiconductors like Si and GaAs, it is far more stable than most known hybrid materials, and some of them even have a shelf life of over one and half decades! More significantly, these findings indicate that the observed structural degradation may not even be intrinsic in nature rather due to some extrinsic mechanisms, because of the variation in shelf life and the non-uniform degradation.

To further examine the microscopic scale crystallinity and how it might be affected by aging, a more detailed comparison in PL and electrical conductivity between S19 and S07 is provided in **Figure 2**. **Figure 2**a depicts their PL spectra in a larger range up to 1200 nm, showing minimal below bandgap PL emission over the entire range, which indicates that there are very few radiative defect centers in the materials. The influence of the potential non-radiative defect centers can be effectively probed by the excitation density dependence of PL intensity over a large range of excitation density.[16, 35] Indeed, as shown in **Figure 2**b, when the excitation density (p) is varied by six orders, both S19 and S07 exhibit a very close to linear dependence (I ∝ $p^n$, n = 0.98 ± 0.01 for S19 and 1.01 ± 0.01 for S07), implying a near 100% internal quantum efficiency, superior than the best reported CdTe, GaAs and hybrid perovskite.[16] Furthermore, S19 and S07 are found to have very similar electron mobility along the *b* axis or vertical direction, 2.5x10$^{-3}$ cm$^2$/(Vs) for S19 and 8.8x10$^{-3}$ cm$^2$/(Vs) for S07, when their I-V characteristics, shown in **Figure 2**c, are fit with the Mott-Gurney model for the space charge limited current.[36] These mobility values are relatively low compared to conventional inorganic semiconductors, which is perhaps understandable for the conductivity along the stacking direction, but nevertheless significantly higher than the mobilities in most organic materials.[37] The approximate linear dependence of J ∝ $V^2$ indicates that the material is highly insulating with minimal free carriers, as expected by the Mott-Gurney law. These results further confirm that the materials have very little either radiative or non-radiative defect centers, thus, very high degree crystallinity in microscopic scale, and may retain the high crystallinity well over a decade.

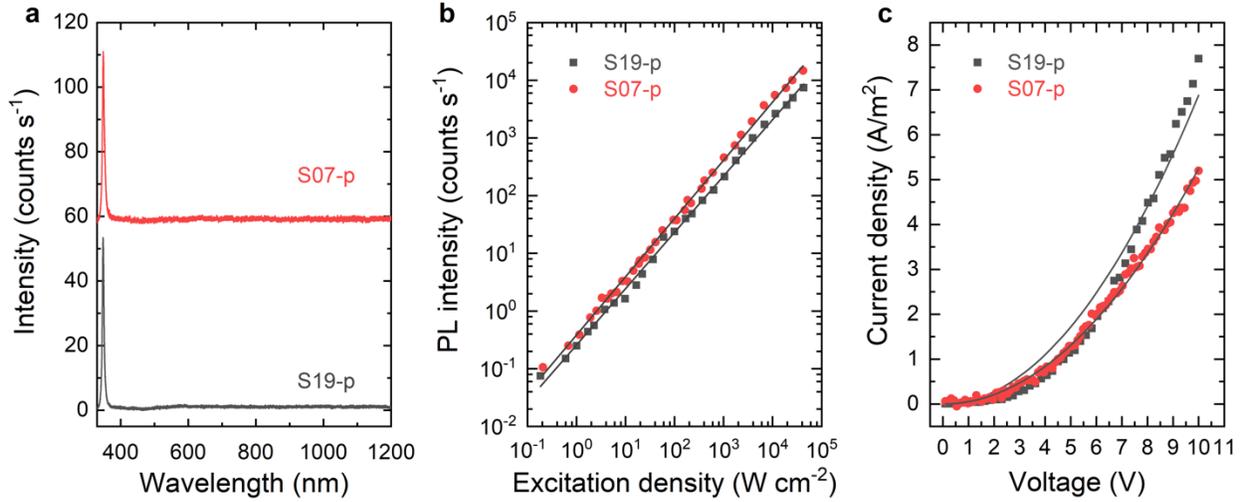

FIG. 2. Comparison between recently synthesized (S19-p) and aged (S07-p) sample in PL and vertical electrical conductivity. (a) Large range PL spectra (S07-p is vertically displaced for clarity), (b) PL intensity *vs.* excitation density, and (c) I-V characteristic (dots – data, lines – fitting curves).

We also note that β-ZnTe(en)$_{0.5}$ exhibits considerably better photo-stability than MAPbI$_3$. In air with prolonged exposure to above bandgap excitation, the former has a degradation threshold ~ 300 W/cm$^2$, and the latter ~ 10 W/cm$^2$,[16] compared to CZTSe ~ 3x10$^4$ W/cm$^2$ and Si above 10$^6$ W/cm$^2$.[38]

## Crystallinity and stability probed by X-ray diffraction

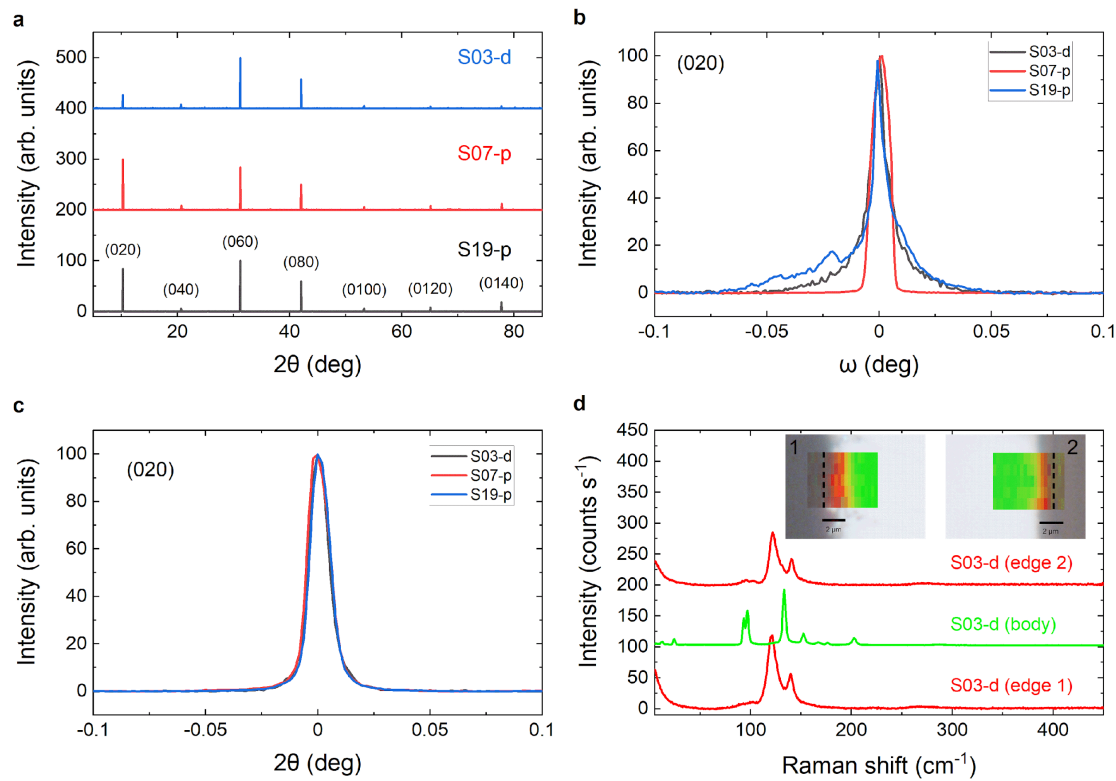

FIG. 3. XRD and Raman characterization of pristine and seemly degraded samples. a) b axis XRD 2θ scans for S19-p, S07-p, and S03-d. b) and c), High-resolution XRD rocking curves or ω scans and 2θ scans, respectively, at (0,2,0) reflection for the three samples. d) Raman spectra of S03-d measured from the cleaved edge at locations close and away from the surfaces (insets: Raman mapping of the Te mode, overlaid with the optical image, on the cleaved edge near the upper and lower surface of the crystal, where the dashed lines indicate the sample surfaces).

It was puzzling that judged by the optical signatures, samples like S06-d were severely degraded, but visually they remained intact. Thus, we performed XRD studies on a few pristine and "degraded" samples used in the optical studies. It turned out that these "degraded" samples still gave rise to the same crystal structure as the pristine sample based on the XRD powder diffraction analyses: for instance, nearly identical (a, b, c) lattice constants between S19-p and S06-d (see detailed comparisons in Table S2).

We further performed HRXRD measurements on selected pristine and "degraded" samples. **Figure 3**a shows very similar (0n0) single crystal diffraction peaks between S19-p, S07-p, and S03-d, except for the (020) peak of S03-d being significantly weaker (see Table S3 for tabulated relative intensities with comparison to simulated results). They all exhibit a very high degree of macroscopic crystallinity, as indicated by the single crystal XRD results of rocking curve or ω scan linewidths ($W_{RC}$) and ω-2θ coupled scan linewidths ($W_{2\theta}$). Roughly speaking, $W_{RC}$ measures the lateral uniformity of the crystalline structure, and $W_{2\theta}$ reflects the uniformity of lattice spacing along the stacking direction. For the (020) diffraction, $W_{RC} = 22''$ and $W_{2\theta} = 38''$ for S19-p, as shown in **Figure 3**b and **3**c, respectively, compared to the results of a high quality ZnTe single crystal with $W_{RC} = 19''$ and $W_{2\theta} = 16.5''$ for the (002) diffraction.[39] For comparison, the best reported $W_{RC}$ is $35''$ for bulk MAPbI$_3$[40] and $30''$ for epitaxially grown GaN.[41] Interestingly, the $W_{RC}$ results of S07-p and S03-d are very similar, $35''$ and $31''$, respectively. The XRD results of S03-d suggest that despite the signs of degradation in optical spectroscopy, the hybrid samples retained the same crystal structure as a whole. Keep in mind that the mm size X-ray beam not only averaged over the entire sample area (typically a few hundred μm) but also the entire thickness (typically 10-20 μm), whereas the spectroscopy measurements were performed with a confocal optical system that only probed a small volume near the sample surface (on the scale of one μm). Raman depth profiling was conducted for the degraded sample using 532 nm laser that is supposed to be transparent to the pristine material, but the Raman signal decayed quickly into the sample, suggesting the surface region of the degraded sample was highly absorptive to this below bandgap wavelength.

We then cleaved a severely degraded crystal S03-d and performed Raman mapping on the newly exposed edge, shown as an inset in **Figure 3**d. It turned out that the body of the about 20 μm thick crystal exhibited the same Raman modes as a pristine sample, and only the regions approximately 2 μm thick from the top and bottom surface were degraded, as shown in **Figure 3**d. The results explain the puzzling "inconsistency" between the optical and XRD results, including the weakened (020) peak of S03-d, corresponding to the largest incidence angle at which more surface region was probed. The observation of the Te modes from the surface region of the "degraded" hybrid is similar to thermally oxidized ZnTe where the top layer consists of amorphous ZnO and nanoscale Te domains.[42] The observations seem to suggest that the degradation of the hybrid is caused by surface oxidation, as to be confirmed next, that occurs in most non-oxide semiconductors. We may further speculate that a terminating layer of en molecules might prevent the surface from oxidation, a self-passivation effect. However, the oxidation process might initiate at a defective surface or a region with structural defects. This understanding is consistent with the observed variation in shelf life and degradation inhomogeneity. We have also noticed that the edges of a newly synthesized crystal tend to degrade faster than the top surface, because the edges are not terminated by en molecules (see **Figure. 1**a).

**Stability probed by surface analyses**

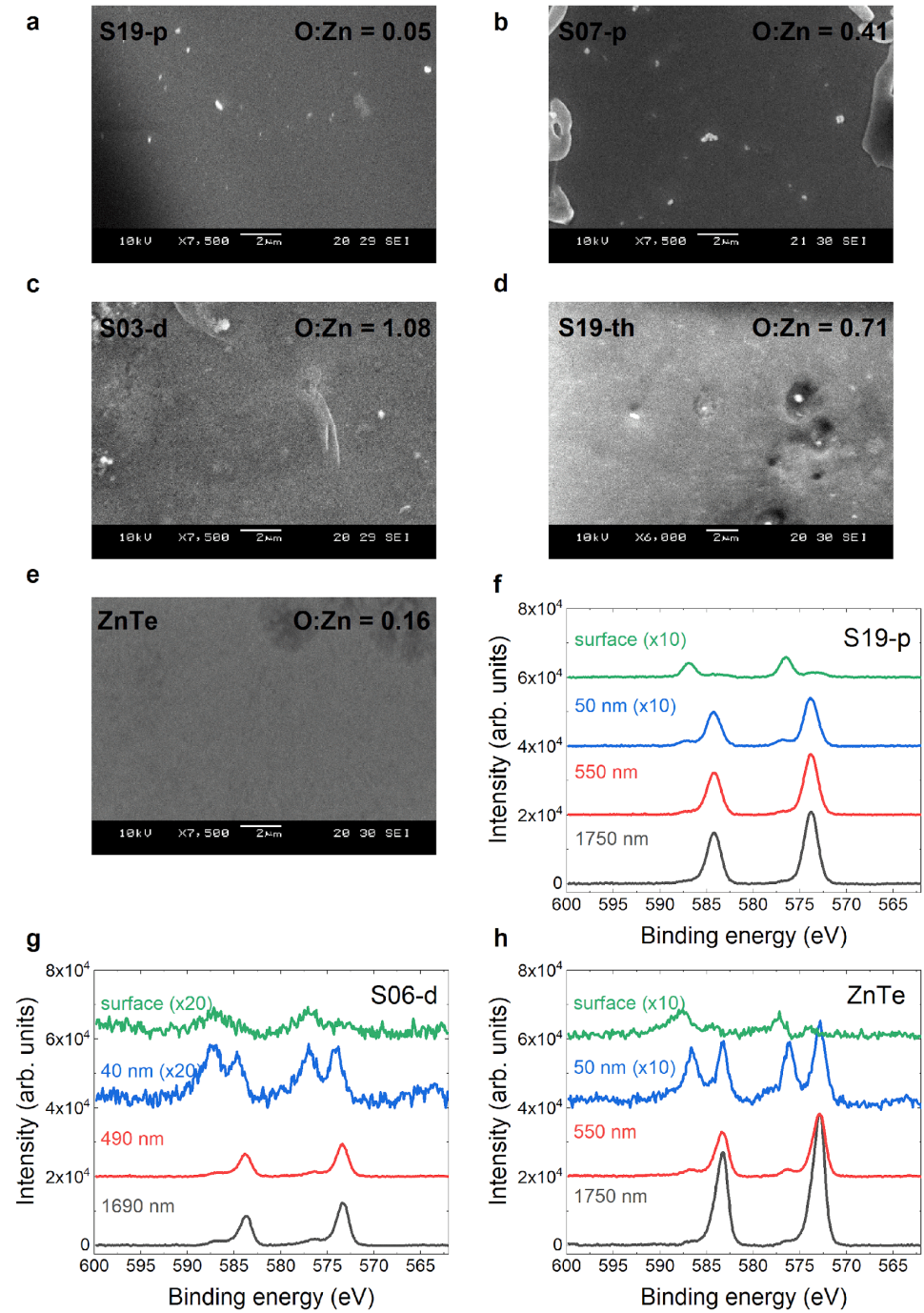

FIG. 4. SEM and XPS surface characterization and analyses. a–e) SEM images of a fresh sample (S19-p) (a), an aged sample showing very little degradation optically (S07-p) (b), an aged sample optically showing severe degradation (S03-d) (c), a thermally degraded fresh sample (S19-th) (d) and single crystalline ZnTe (e). (f–h) XPS spectra at different etching times or depths of sample S19-p (f), sample S06-d (g) and ZnTe (h).

To verify the oxidation process and degradation mechanism, we performed surface analyses using scanning electron microscopy (SEM), energy-dispersive X-ray (EDX) analysis, and X-ray photoelectron spectroscopy (XPS) on multiple pristine and "degraded" samples. **Figure 4**a–e shows the SEM images obtained for S19-p, S07-p, S03-d, and S19-th (thermally oxidized at 190 ºC for 20 minutes), and ZnTe, respectively, taken from smooth and clean areas of the samples where optical measurements were conducted. The O:Zn ratio obtained from the EDX analysis is shown with each SEM image. The original EDX spectra are depicted in **Figure S1**. The comparison reveals a steady increase in oxygen content on the surface with increasing level of degradation, for instance, O:Zn = 0.05 for S19-p and 1.08 for S03-d. These results qualitatively support the proposed degradation mechanism of surface oxidation.

XPS data further corroborate the above results. We measured depth-resolved XPS spectra for S19-p, S06-d, and ZnTe by applying *in-situ* Ar etching, with the results shown, respectively, in **Figures 4**f-h near the energies of two Te 3d transitions. Additional scans of a broader energy range can be found in **Figure S2**. For S19-p, on the non-etched surface, two peaks were at 576.5 eV ($3d_{5/2}$) and 586.9 eV ($3d_{3/2}$), similar to a naturally oxidized ZnTe in air as reported in the literature;[43] after etching, two new peaks emerged on the lower energy sides respectively at 573.8 eV ($\delta E = -2.7$ eV) and 584.2 eV ($\delta E = -2.7$ eV), and after about 15 second etching (or removal of around 150 nm top layer), the high energy peaks disappeared and the new peaks reached saturated intensities. Similar shifts were observed in ZnTe, but the transition energies for pure ZnTe are slightly lower than in the hybrid, by about 0.8 eV. The difference can be understood by the fact that the Te bonding situations are different in the two structures: with each Te bonding to only three Zn in the hybrid and four Zn in ZnTe. For the severely degraded sample, a similar trend was observed, but a much longer etching time (170 s, approximately corresponding to 1.7 µm in depth) was required to reach a steady state, which suggests a thicker oxidized layer on this severely degraded sample and is consistent with the cleaved edge Raman results (**Figure 3**d). The remaining weak oxidation peaks after etching indicates that oxidation might have somewhat penetrated into the deeper volume of the crystal. Note that similar to XRD, XPS probes a large area due to its macroscopic beam size, therefore, the results could be affected by the embedded extended defects and surface contamination, whereas micro-Raman and EDX are able to analyze visually perfect regions. We note that even the new sample S19-p had a thin oxide layer, although showing no effect in optical spectroscopy after being kept in air for one year. The existence of the pristine area in S03-m seems to suggest that the oxidation process is self-limiting, if no structural defects, as in most non-oxide inorganic semiconductors.

**Thermal stability in different environments**

Thermal stability is important for device application because of a self-heating effect under the operation condition. Thermal stability study was performed in both air and $N_2$ protected condition on newly made crystals (S19-p) in two ways: (1) By heating one single piece of the crystal using a heating stage with a small heating chamber from 40 to 500 ºC while monitoring with Raman spectroscopy *in-situ*. (2) By performing thermogravimetric analysis (TGA) to a collection of small crystals. These two experiments are similar in the sense that in both cases a material property is probed when the sample temperature is raised in a small step and held steady briefly at each step to take the measurement, but in the former case a Raman spectrum is measured *versus* the weight in the latter case.

In Raman, a significant difference in thermal degradation was observed between heating in air or $N_2$ until approximately 150 ºC. **Figure 5**a shows the evolution of the Raman spectra while heated in air (spectra at more temperature points can be found in **Figure S3**), where the Te peaks start to grow at around 160 °C, and the hybrid peaks vanish above 210 °C, mimicking an accelerated degradation process in ambient condition. **Figure 5**b shows the results under $N_2$ protection (more spectra in **Figure S3**). In this case, no or very little Te related Raman signal can be observed for temperature up to 300 °C; but at around 230 °C, ZnTe LO phonon starts to emerge (**Figure S3**). When brought back to room temperature after heated up to 500 °C, the sample exhibits the characteristic multiple LO phonon resonant Raman modes of a somewhat defective ZnTe [39] (**Figure 5**c), which is similar to the report where thermal annealing of ZnS(en)$_{0.5}$ at 600 °C in vacuum yielded ZnS.[44] However, if heated in air, no ZnTe is found in the degradation product either at high temperature (up to 400 °C, **Figure 5**a) or after returning to room temperature (**Figure 5**c) where instead a mixture of Te and $TeO_2$ is seen.[45] Thus, $N_2$ to a large extent prevents the oxidation process, and the process reflects the intrinsic degradation path: evaporation of en molecules. The finding provides guidance to the acceptable operating temperature of a potential device using this material with proper encapsulation to minimize the oxidation.

The TGA results, shown in **Figure 5**d, are qualitatively consistent with the spectroscopy results but offer more insight. Initially till around 230 °C, both curves show slow mass reduction, but the curve in $N_2$ is slightly faster (inset of **Figure 5**d), which can be explained by that while ligand evaporation or calcination occurs due to heating in both cases, in air the mass loss is made up partially by oxidation. The TGA curve in $N_2$ shows continual decrease but accelerated near 290 °C, reaching a plateau before 400 °C, corresponding to the complete conversion into ZnTe, which matches the mass loss of the molecules (about 13% in weight). In contrast, in air, significant mass loss does not occur until around 200 °C, and at about 290 °C, the curve actually begins to increase, reflecting the oxidation process is accelerated; and the mass gain saturates at around 600 °C. We may understand the intriguing curve as a convoluted effect of multiple competing oxidation processes: $ZnTe(en)_{0.5} \rightarrow ZnTeO_2$, where $ZnTeO_2$ is predominantly a mixture of $ZnO + xTe + yTeO + zTeO_2$ (x + y + z =1), where the highest point matches $ZnO + 0.5Te + 0.5TeO_2$. The detailed analyses are given in **Figure S4**, although the exact degradation process requires further study.

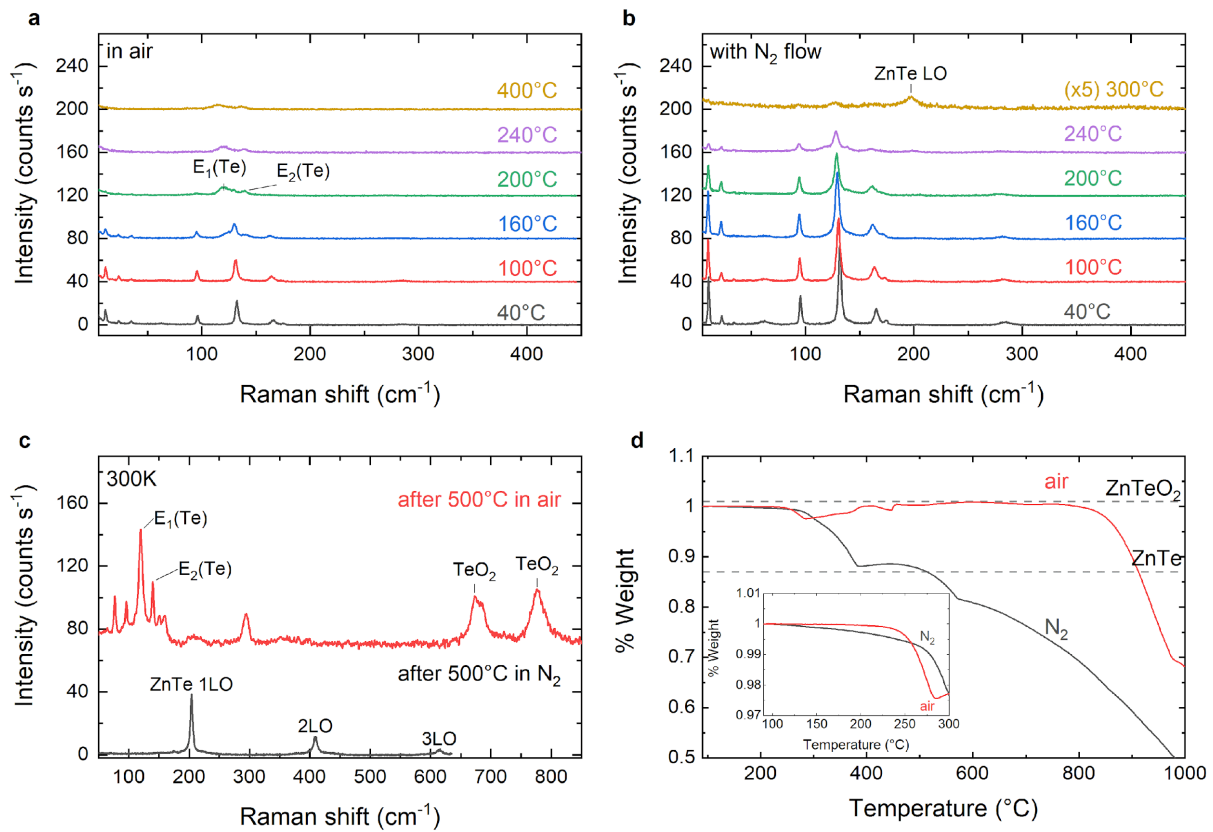

FIG. 5. Thermal degradation of freshly made sample (S19-p) in air and N2 environment probed by Raman and thermogravimetric analysis (TGA). a) *In-situ* Raman spectroscopy in air. b) *In-situ* Raman spectroscopy in N2. c) Raman spectra at room temperature after being heated to 500 °C in air and N2 then returning to room temperature. d) TGA in air and N2 (inset: low temperature region).

We may have a unified picture for the thermal degradation: following the initial loss of en near the surface, the molecules in the body start to leave at around 290 °C, and at close to 400 °C, the hybrid is fully degraded, resulting in different products in different environments.

Although naturally occurring slow degradation can be mimicked to some extent by an accelerated process – heating in air, heating process is not able to reveal some subtle, often extrinsic, effects that actually happened in the real process, such as the region to region or sample to sample variations. It is also non-trivial to establish a reliable quantitative correlation of the thermal degradation temperature with the shelf life of the material other than the qualitative correlation. This work has provided one correlated study.

**Formation energies: DFT calculation and experiment**

The formation energy of β-ZnTe(en)$_{0.5}$ was calculated by $\Delta E = E[\beta\text{-ZnTe(en)}_{0.5}] - E[\text{ZnTe}] - \frac{1}{2} E[\text{C}_2\text{N}_2\text{H}_8]$, which is relevant to the intrinsic degradation path mentioned above, where $E[\beta\text{-ZnTe(en)}_{0.5}]$ is the total energy per formula unit of β-ZnTe(en)$_{0.5}$, $E[\text{ZnTe}]$ is the total energy per formula unit of ZnTe in zinc-blende structure and $E[\text{C}_2\text{N}_2\text{H}_8]$ is the total energy of $\text{C}_2\text{N}_2\text{H}_8$ molecule. The calculated formation energy is the dissociation energy without including the kinetic barrier. The calculation has yielded $\Delta E$ = -0.52 eV per formula, compared to -0.80 eV of the formation energy of ZnTe (typical for II-VI compounds), and is much greater than that of (CH$_3$NH$_3$)PbI$_3$ in the range of ±0.1 eV.[13, 14] Formation energy is the lower bound of the thermal activation energy of the intrinsic thermodynamic degradation process. The kinetic barrier between the degradation products and the hybrid state is also relevant, particularly when the formation energy is very small, such as in the case of the hybrid perovskite where the kinetic barrier plays a critical role in stabilizing the structure.[46] The situation resembles the chemical reaction described by a well-known transition state theory. A sufficiently large formation energy may be viewed as a necessary condition for the desired level of long-term stability, because it describes the thermodynamics stability in vacuum or protected environment. A thermodynamically stable structure may still be environmentally unstable, such as silicene which is stable in vacuum but usually oxidized very quickly in air.[47] The observed long-term stability in part of S03-m and in S07-p implies that surface oxidation can be self-limited under favorable growth conditions, for instance, when the surface is uniformly passivated by the organic molecules.

To gain further insight into the potential intrinsic long-term stability, we have measured the degradation time of β-ZnTe(en)$_{0.5}$ in elevated temperatures between 210 – 270 °C by monitoring the 133 cm$^{-1}$ Raman mode intensity *vs.* time at each set temperature. The Raman data of Fig. 5(b) could be viewed as the results at time t ≈ 0. The intensity exhibits exponential decay with time a temperature dependent decay rate. The degradation rate may be defined as k(T) = 1/t$_{1/2}$, where t$_{1/2}$ is the half intensity point. Assuming k(T) = A exp(-E$_A$/kT), fitting the experimental data has yielded E$_A$ = 1.63 ± 0.06 eV. The details can be found in **Figure S5**. The difference between the measured activation energy and the calculated formation energy could be explained as the kinetic energy barrier height (approximately 1.11 eV). This study suggests that the intrinsic shelf-life (from the extrapolated t$_{1/2}$ at 25 °C in N$_2$) of this material could be as long as 1.44 x 10$^8$ years.

**Further discussion**

In general, the stability includes three key aspects: (1) Intrinsic effect, namely the material formation energy calculated as an infinite crystal, which is expected to be smaller for a hybrid material than its inorganic component. Thus, the consideration lies in whether the binding strength of the hybrid material is adequate for the intended application condition (*e.g.*, operation temperature). (2) The surface or edge effect, which is most relevant to the chemical stability of the material, where degradation can initiate through processes such as oxidation and/or evaporation of the terminating molecules. (3) The structural defects, which may provide easier paths for degradation. The substantial variation in shelf life from sample to sample even between different regions within one sample indicates that the observed degradation might not be intrinsic in nature, but be associated with surface imperfection or structural defects. Non-uniform degradation often occurs in inorganic semiconductors due to the spatial variation in defect condition, for instance, in GaAs[48] and CZTSe.[38] The detailed degradation paths of the hybrids under different environments require further studies. The most aged sample (S03-m) still showing the pristine spectroscopy features suggests the lower bound of the thermodynamically determined material lifetime to be around 16 years, although the thermal degradation study indicates that it could be much longer.

$\beta$-ZnTe(en)$_{0.5}$ can be viewed as one of the most perfect man-made superlattices and the only hybrid superlattice that has been carefully investigated. It also represents a structure with unusual chemical bonding (*i.e.*, 3-fold bonded Te atoms) yet without the detrimental effects of dangling bonds. Despite the structure complexity, a hybrid material can have a very high degree of crystallinity, manifested in both macroscopic structural properties (*e.g.*, XRD linewidth) and microscopic properties (*e.g.*, defect emission). This attribute can offer much-needed quantum coherence for electronic processes in next-generation (opto)electronic technologies. This and other related II-VI based hybrids demonstrate a totally different approach for forming a perfectly abrupt heterostructure, that is, combining two structures with drastically different materials, instead of two structurally similar ones, to eliminate the intermixing that occurs in most conventional heterostructures. An inorganic analog is the abrupt interface between rock-salt PbTe and zinc-blende CdTe along their [111] direction.[49]

**Conclusions**

By providing a comprehensive characterization of one prototype system, β-ZnTe(en)$_{0.5}$, with some samples being monitored over 16 years, this work illustrates the interplay of intrinsic and extrinsic degradation mechanisms in determining the long-term stability of an organic-inorganic hybrid material. Benefiting from its relatively large intrinsic formation energy as well as a large kinetic barrier, and stable surface, even without encapsulation, β-ZnTe(en)$_{0.5}$ has been shown to exhibit over 15-year shelf life, whereas its intrinsic lifetime could be as long as $10^8$ years. The observed deterioration of structural integrity is actually caused primarily by the extrinsic effects, such as surface imperfection, exposed edge, and structural defects. This study indicates that formation energy can serve as an effective screening parameter for the long-term stability prospective of a new hybrid material. However, when the formation energy is adequately high, extrinsic degradation paths could be practically more significant for the long-term stability of the hybrid.

β-ZnTe(en)$_{0.5}$ also exhibits exceptionally high degree of both macroscopic and microscopic scale structural perfectness, manifesting as small XRD and Raman linewidths comparable to the high quality ZnTe and near 100 % internal PL quantum efficiency superior than most known high bandgap inorganic semiconductors such as ZnO and GaN.

The success in the synthesis of these practically perfect hybrid superlattices demonstrates a nonconventional strategy to achieve periodic stacking of hetero-structure materials with abrupt interfaces and offer a more practical method to controllably stack ultra-thin 2D layers to achieve desirable overall thickness to meet a specific application need.

      This study provides practical guidance for selecting different organic-inorganic hybrid materials to suit different application needs, and reveals the potentials of the II-VI based hybrids for a wide range of scientific explorations and applications with excellent long-term stability.

**Materials and Methods**

*Materials*: New samples of β-ZnTe(en)$_{0.5}$ were synthesized following slight modifications of previously reported literature procedures.[25] ZnCl$_2$ (333 mg, 2 mmol), Te (128 mg, 1 mmol), and ethylenediamine (4 mL) were placed in a Teflon-lined stainless steel high-pressure acid digestion vessel (Parr model 4746). The vessel was placed in a muffle furnace set at 200 °C for 10 days. After slow cooling to room temperature, the resulting product mixture was filtered and washed with distilled water, 95% ethanol, and diethyl ether, then dried in air to produce colorless plate-like crystals. The samples used in this work are labeled as Sxx-y, where xx indicates the year of synthesis (*e.g.*, 03 for 2003), and y indicates the degradation status (*e.g.*, p for pristine, d for degraded). Samples labeled as S19-p were from one newly synthesized batch. The oldest samples (2006 or earlier) were the same pieces mounted on their sample holders used in previous studies,[25-27, 50] and the 2007 samples were from one synthesis batch and were not studied previously. Majority of S07 showed no sign of degradation, with few exceptions (*e.g.*, S07-pd). The aged samples were stored under ambient condition until they were (re-)measured recently.

*Optical characterization:* Raman and PL measurements were carried out on a Horiba LabRam HR800 confocal Raman microscope with a 1200g/mm grating. For Raman, a 532 nm laser beam was focused on the sample surface using a long working distance 50x microscope lens with NA = 0.5. The laser power focused onto the sample was 125 ± 10 µW (~9.4x10$^3$ W/cm$^2$). For PL, the sample was excited by a 325 nm laser beam, which was focused through a 40x UV microscope lens with NA = 0.5. The laser power delivered to the sample was 15.5 µW (~3.1x10$^3$ W/cm$^2$).

*Electrical characterization:* Vertical conductivity measurements were performed for S19-p and S07-p. The I-V scan was done by using a Keithley 2401 Source Meter Unit using Keithley Kickstart software. The sample with surface normal along the b axis was placed on a chip carrier that provided the bottom contact. Tungsten probe tip (of 50 μm in diameter) was placed directly on the sample surface as the top contact. Multiple samples were measured for each batch of the samples with fluctuating results likely due to inconsistency in contacts. The data shown represent approximately the best achieved results for the respective batches of the samples. The I-V characteristic was fit to Moo-Gurney law:[36] $J = 9/8\ \varepsilon\varepsilon_0\mu V^2/L^3$, where J is the current density, $\varepsilon \approx 6$ is the dielectric constant of β-ZnTe(en)$_{0.5}$, μ is the carrier mobility, and L is the sample thickness (L = 6 and 10 μm, respectively for S19-p and S07-p with about 10% accuracy).

*Structural characterization:* X-ray crystallography data were acquired with an Agilent (now Rigaku) Gemini A Ultra diffractometer. Crystals of suitable size were coated with a thin layer of paratone-N oil, mounted on the diffractometer, and flash cooled to 105 K in the cold stream of the Cryojet XL liquid nitrogen cooling device (Oxford Instruments) attached to the diffractometer. The diffractometer was equipped with sealed-tube long fine focus X-ray sources with Mo target (λ = 0.71073 Å) and Cu target (λ = 1.5418 Å), four-circle kappa goniometer, and CCD detector. CrysAlisPro software was used to control the diffractometer and perform data reduction. The crystal structure was solved with SHELXS. All non-hydrogen atoms appeared in the E-map of the correct solution. Alternate cycles of model-building in Olex2 and refinement in SHELXL followed. All non-hydrogen atoms were refined anisotropically. All hydrogen atom positions were calculated based on idealized geometry and recalculated after each cycle of least squares. During refinement, hydrogen atom – parent atom vectors were held fixed (riding motion constraint). High resolution X-ray diffraction (HRXRD) measurements were performed at room temperature on an XRD diffractometer (PANalytical's X'Pert PRO) with triple-axis configuration using monochromatized Cu Kα radiation (1.5418 Å). To compare the line shape of 2θ-ω coupled scans and rocking curves, peak intensities were normalized, and the peak positions were centered at zero.

*Surface analysis:* The morphology of the hybrid samples was investigated by using scanning electron microscope (JEOL JSM-6480) equipped with x-ray energy dispersive spectrometry (EDX). EDX spectra were obtained with 10 keV beam energy. X-ray photoelectron spectroscopy (XPS; Thermo Scientific ESCALAB XI+) was carried out using 200 μm diameter monochromatized Al source (hv = 1486.6 eV). For depth profiling, sample surface was repeatedly sputtered with an argon beam (3000 eV, 500 μm diameter) and analyzed by XPS. The etch rate was estimated to be 10 nm/s. Binding energies were calibrated relative to the C 1s peak at 284.8 eV.

*Thermal study:* Temperature-dependent Raman measurements were carried out with a heating system Linkam TS1500. At each temperature step, five minutes were allowed for thermal stabilization of the sample before Raman spectrum was measured. A 10 ºC/minute ramping rate was used to heat the sample to the next temperature step. $N_2$ protected condition was achieved by a mild sustained $N_2$ flow into the chamber during the entire temperature profile.[51] Thermal degradation results shown in Fig. S5 were obtained in the same way under the $N_2$ condition with a ramping rate of 20 ºC/minute to reach the targeted temperature and by taking a Raman spectrum every 5 minutes. Thermogravimetric analysis was performed using Mettler Toledo TGA/SDTA851 instrument (Mettler-Toledo AG Analytical, Schwersenbach, Switzerland) in both air and $N_2$ surroundings. Blanks were run under identical conditions for each gas and subtracted from the respective TGA curves to correct for buoyancy. Ramping rate was 5 ºC/minute.

*Density-functional theory modeling:* Density functional theory (DFT) calculations were carried out using the Vienna *ab initio* simulation package (VASP).[52] The projector augmented wave potentials were used to describe the interaction between the ion-cores and valence electrons. The PBEsol generalized gradient approximation was used for the exchange–correlation functional. Plane-waves with cutoff energy of 50 Ry were used as basis set. The formation energy of β-ZnTe(en)$_{0.5}$ was calculated by taking the total energy difference between bulk β-ZnTe(en)$_{0.5}$ and the sum of zinc-blende ZnTe and $C_2N_2H_8$ molecule in the corresponding stoichiometry. ZnTe was calculated in zinc-blende structure and $C_2N_2H_8$ molecule was calculated in a cubic supercell with an edge length of 15 Å. The accuracy of the formation energy calculation is around 10%.[53] Raman spectrum calculations (frequency and symmetry) were carried out following Ref..[54]

**Acknowledgement**

We thank Dr. Kristen Dellinger of UNCG for assistance in XPS measurements, Dr. Jianhua Li of Rice Univ. and Zahirul Islam of ANL for single-crystal XRD measurements, Dr. Haitao Zhang of UNCC in SEM/EDX measurements, and Dr. Brian Fluegel of NREL for locating some aged samples. The work at UNCC was supported by ARO/Physical Properties of Materials (Grant No. W911NF-18-1-0079), University of North Carolina's Research Opportunities Initiative (UNC ROI) through Center of Hybrid Materials Enabled Electronic Technology, and Bissell Distinguished Professorship. VESTA 3 was used to produce drawings of crystal structures.

**Supporting Information Available Online:** EDX and atomic ratio analysis; XPS spectra at different etching times or depths in an extended spectral range; *in-situ* heating studies using Raman spectroscopy for a freshly made sample S19-p in air and $N_2$ at more temperature points; analyses of TGA curves for S19-p in air and $N_2$; estimation of activation energy $E_A$ of thermal degradation by monitoring the intensity decay of the 133 cm$^{-1}$ Raman mode of the hybrid structure at different elevated temperatures; observed Raman mode frequencies and the corresponding calculated phonon mode frequencies and their symmetry assignments; powder diffraction analyses of S19-p and an optically degraded sample S06-d; measured and simulated relative intensities for XRD (0,n,0) reflection peaks.

**Associated Content**

This paper has been submitted to preprint server: Tang Ye; Margaret Kocherga; Yi-Yang Sun; Andrei Nesmelov; Fan Zhang; Xiao-Ying Huang; Jing Li; Damian Beasock; Daniel S. Jones; Thomas A. Schmedake; Yong Zhang. Highly Ordered II-VI Based Organic-Inorganic Hybrids with Over 15-Year Shelf Life. 2020, arXiv:2006.00349v2. arXiv.org. https://arxiv.org/abs/2006.00349 (submitted on 30 May 2020).

**For Table of Contents Only**

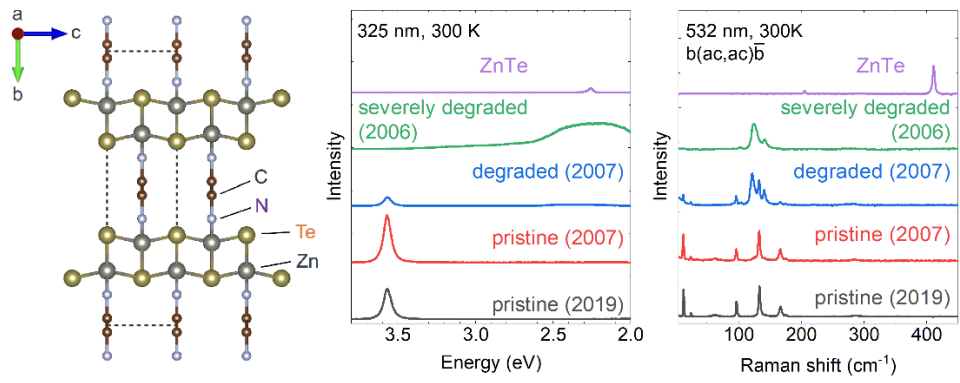

**Supporting Information**

# II-VI Organic-Inorganic Hybrid Nanostructures with Greatly Enhanced Optoelectronic Properties, Perfectly Ordered Structures, and Over 15-Year Shelf Stability

*Tang Ye[1], Margaret Kocherga[1], Yi-Yang Sun[2], Andrei Nesmelov[3], Fan Zhang[4], Wanseok Oh[4], Xiao-Ying Huang[5,6], Jing Li[5], Damian Beasock[1], Daniel S. Jones[3], Thomas A. Schmedake[1,3], and Yong Zhang[1,4]\**

[1]Nanoscale Science, University of North Carolina at Charlotte, Charlotte, NC 28223, USA.

[2]State Key Laboratory of High-Performance Ceramics and Superfine Microstructure, Shanghai Institute of Ceramics, Chinese Academy of Sciences, Shanghai 201899, China.

[3]Department of Chemistry, University of North Carolina at Charlotte, Charlotte, NC 28223, USA.

[4]Department of Electrical and Computer Engineering, University of North Carolina at Charlotte, Charlotte, NC 28223, USA.

[5]Department of Chemistry and Chemical Biology, Rutgers University, Piscataway, NJ 08854, USA.

[6]State Key Laboratory of Structural Chemistry, Fujian Institute of Research on the Structure of Matter, Chinese Academy of Sciences, Fuzhou, Fujian 350002, P. R. China.

**Figure S1**

a

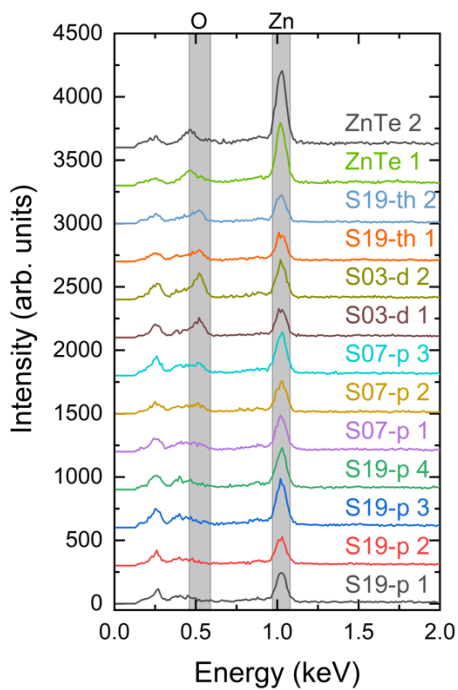

b

|        | Mean | Std error |
|--------|------|-----------|
| ZnTe   | 0.16 | 0.04      |
| S19-p  | 0.05 | 0.02      |
| S07-p  | 0.41 | 0.04      |
| S03-d  | 1.08 | 0.07      |
| S19-th | 0.71 | 0.11      |

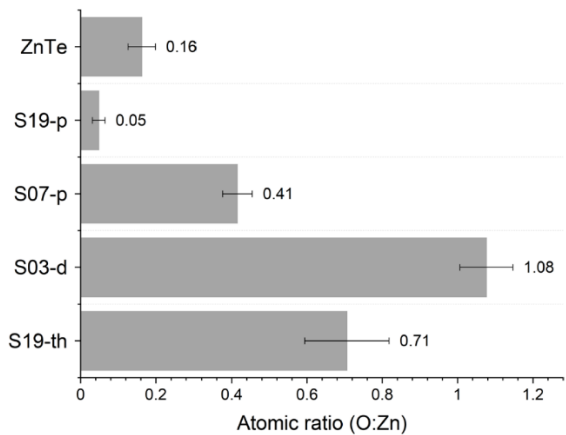

**Figure S1.** EDX and atomic ratio analysis. a) EDX spectra of different samples. b) Atomic ratios of O:Zn obtained from integrated peak intensities of EDX spectra.

**Figure S2**

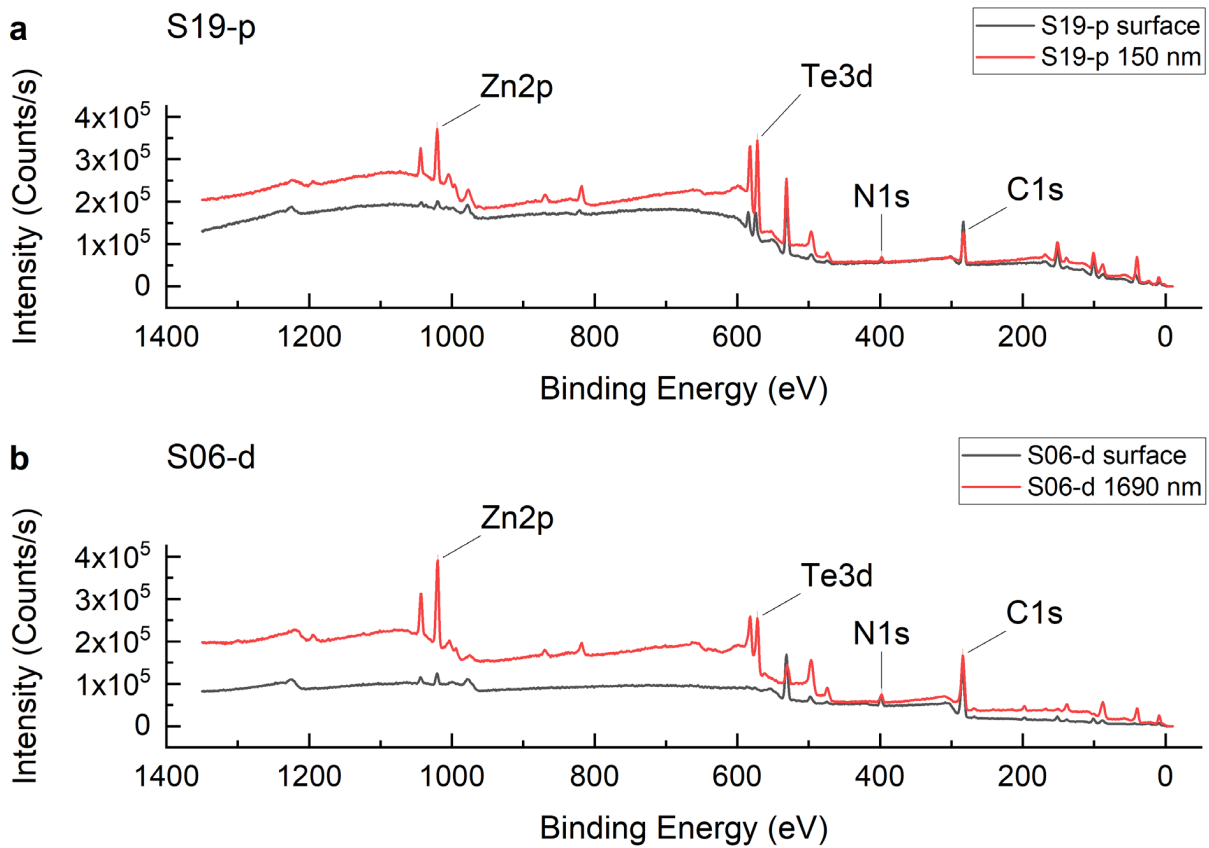

**Figure S2.** XPS spectra at different etching times or depths in an extended spectral range. a) a pristine sample S19-p, b) an optically degraded sample S06-d.

**Figure S3**

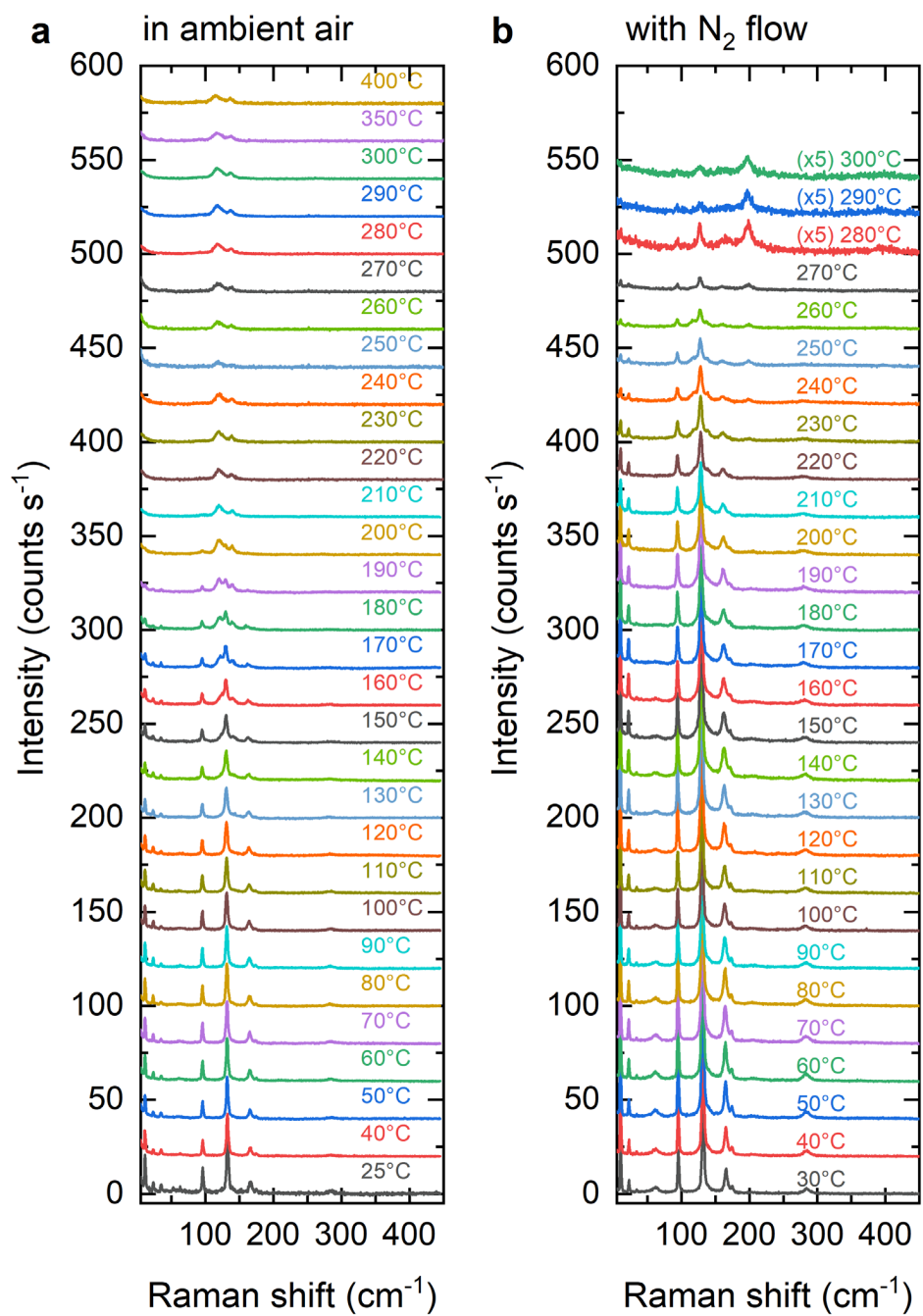

**Figure S3.** *In-situ* heating studies using Raman spectroscopy for a freshly made sample S19-p in air (a) and $N_2$ (b).

Figure S4

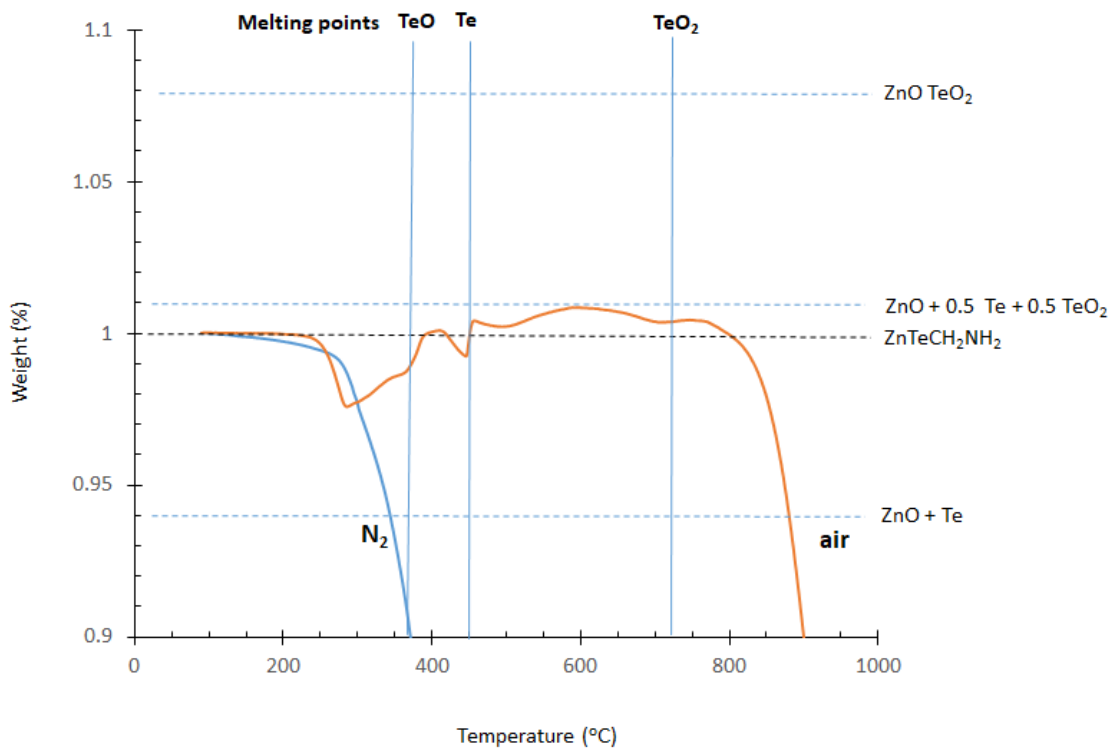

| Formula | MW | Relative mass |
|---|---|---|
| ZnTe(CH$_2$NH$_2$) | 223.0296 | 1.00 |
| ZnTeO$_3$ | 240.9782 | 1.08 |
| ZnTeO$_2$ | 224.9788 | 1.01 |
| ZnTeO | 208.9794 | 0.94 |
| ZnTe | 192.98 | 0.87 |

**Figure S4.** Analyses of TGA curves for a freshly made sample S19-p in air and N$_2$. Horizontal lines indicate the masses of different potential degradation products, with the values show in the table below. The vertical lines indicate the melting temperatures for different Te products: 370, 450, and 720 °C for TeO, Te, and TeO$_2$, respectively [1].

**Figure S5**

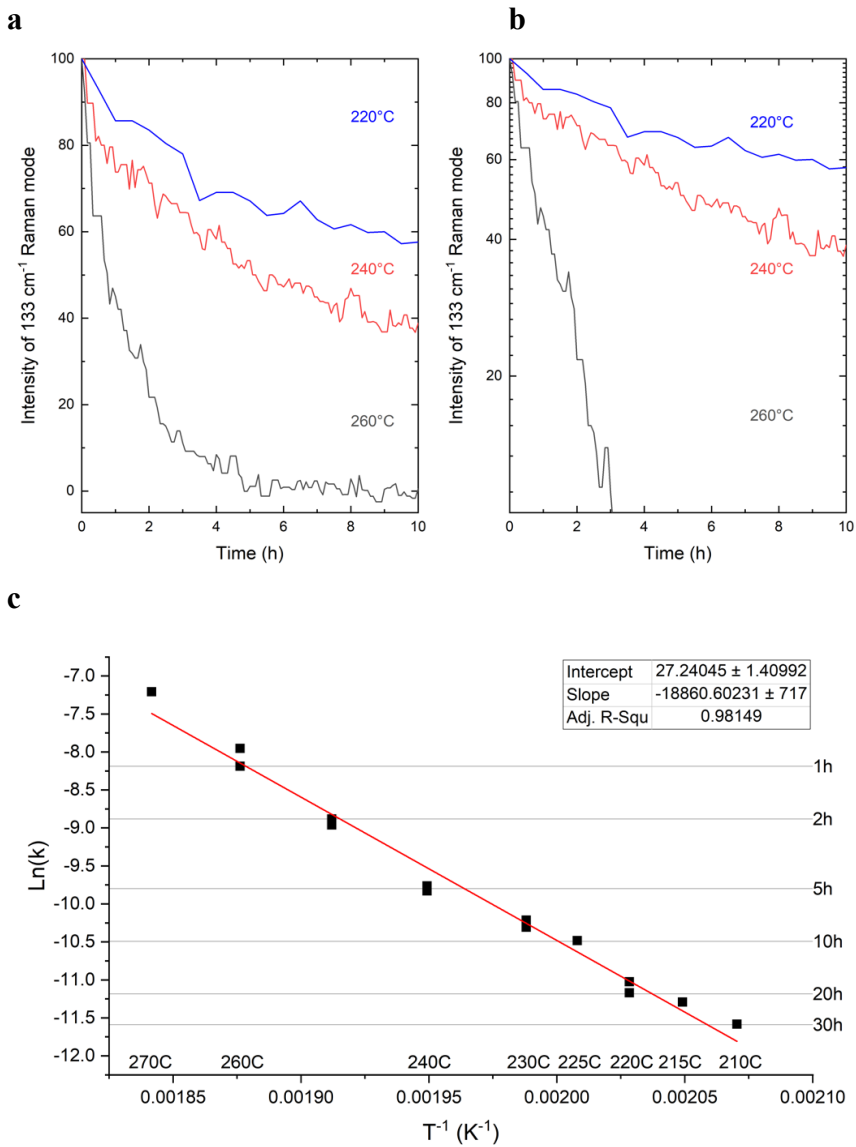

**Figure S5.** Use Arrhenius relationship k = Aexp(-$E_A$/kT) to estimate activation energy $E_A$ of thermal degradation by monitoring the intensity decay of the 133 cm$^{-1}$ Raman mode of the hybrid structure at different elevated temperatures in $N_2$. (a) and (b): peak intensity *vs.* time plots at selected temperatures in linear scale (a), and log scale (b). (c) Linear fit of Ln(k) *vs.* Temperature. The rate constant k at each temperature is determined by 1/$t_{1/2}$, where $t_{1/2}$ is the intensity decay time of the 133 cm$^{-1}$ Raman mode reaching 50% of the t = 0 intensity. The fitting yielded a slope of 18860 ±717 K, corresponding to an estimated $E_A$ = 1.63 ± 0.06 eV; and Ln(A) = 27.24 ± 1.41, corresponding to A = 6.8 (−5.1,+20.9) x 10$^{11}$ s$^{-1}$. Using the fitted line, the estimated $t_{1/2}$(25°C) = 1.44 x 10$^8$ years.

**Table S1**

Observed Raman mode frequencies and the corresponding calculated phonon mode frequencies and their symmetry assignments.

| Figure | Observed frequency [cm$^{-1}$] | Calculated phonon mode frequency [cm$^{-1}$] | Symmetry | Corresponding en mode |
|---|---|---|---|---|
| 1c | 12.11 | 14.34 | B2g | |
| 1c | 24.26 | 22.96 | Ag | |
| 1c | 62.61 | | | |
| 1c | 96.88 | 95.92 | Ag | |
| 1c | 133.38 | 137.12 | Ag | |
| 1c | 167.09 | 172.61 | B2g | |
| 1c | 175.22 | 180.05 | Ag | |
| 1c | 286.74 | 289.71 | Ag | |
| 1d | 540.71 | 538.79 | Ag | |
| 1d | 567.03 | 606.56 | B2g | |
| 1d | 988.53 | 996.32 | Ag | |
| 1d | 1170.35 | 1156.56 | Ag | |
| 1d | 1596.71 | 1579.37 | Ag | |
| 1d | 2868.82 | 2928.87 | Ag | C-H vibration (2860 cm$^{-1}$) |
| 1d | 2916.37 | 2965.59 | B2g | C-H vibration (2928 cm$^{-1}$) |
| 1d | 3128.44 | 3179.35 | Ag | |
| 1d | 3203.68 | 3208.25 | B2g | |

**Table S2**

Powder diffraction analyses of a pristine sample S19-p and an optically degraded sample S06-d, yielding nearly identical (*a*, *b*, *c*) lattice constants.

|  | **S19-p** | **S06-d** | *Ref. (13)* | *Ref. (17)* |
|---|---|---|---|---|
| *a* (Å) | 5.6709(9) | 5.6736(6) | 5.660(1) | 5.6787(2) |
| *b* (Å) | 17.160(3) | 17.169(2) | 17.156(3) | 17.1998(6) |
| *c* (Å) | 4.3403(6) | 4.3445(5) | 4.336(1) | 4.3523(1) |
| *α,β,γ* | 90° | 90° | 90° | 90° |
| Volume (Å³) | 422.37(12) | 423.21(9) | 421.04(14) | 425.10(4) |
| *R1, obs.* | 0.0371 | 0.0365 | 0.0495 | |
| *wR2, obs.* | 0.0766 | 0.0896 | 0.0816 | |

**Table S3**

XRD (0,n,0) reflection peak relative intensities, normalized to (0,6,0). Simulated intensities were calculated from S19-p single-crystal structure data (using Mercury 3.10.2, CCDC).

| h | k | l | S19-p | S07-p | S03-d | Simulated |
|---|---|---|-------|-------|-------|-----------|
| 0 | 2 | 0 | 84.2 | 119 | 27.2 | 75.1 |
| 0 | 4 | 0 | 5.72 | 10.4 | 8.02 | 7.02 |
| 0 | 6 | 0 | 100 | 100 | 100 | 100 |
| 0 | 8 | 0 | 59.6 | 59.4 | 57.4 | 63.7 |
| 0 | 10 | 0 | 5.72 | 7.29 | 5.56 | 6.82 |
| 0 | 12 | 0 | 7.69 | 9.38 | 3.7 | 6.41 |
| 0 | 14 | 0 | 18.3 | 14.6 | 4.94 | 16.3 |

Simulated XRD spectrum of (0,n,0) reflections:

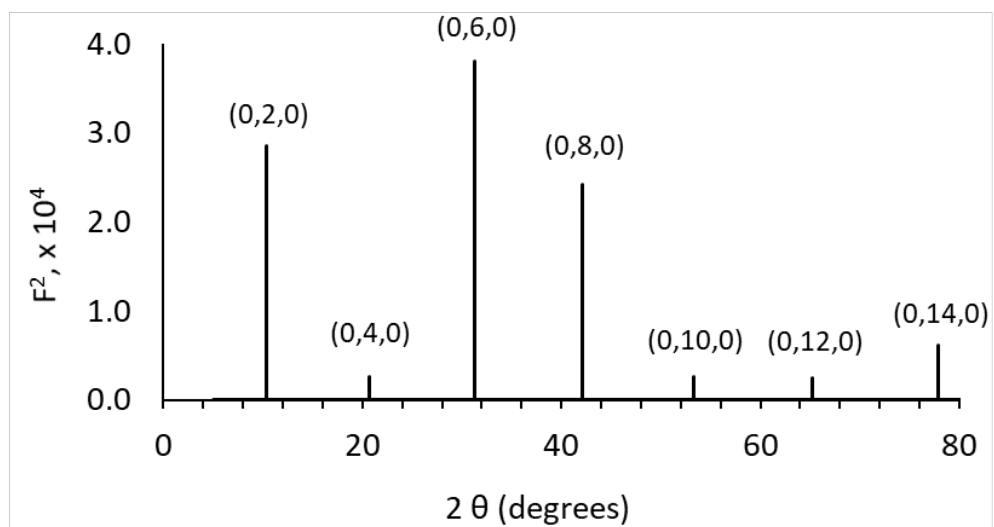